\documentclass[epjCONF]{svjour}
\usepackage{graphics}
\usepackage[varg]{txfonts}
\usepackage[latin1]{inputenc}

\session-title{The Space Photometry Revolution}

\begin{document}

\title{The space photometry revolution and our understanding of RR Lyrae stars}
\author{R. Szab\'o\inst{1}\fnmsep\thanks{\email{rszabo@konkoly.hu}} \and 
J.~M. Benk\H o\inst{1} \and 
	M. Papar\'o\inst{1}\and	
	E. Chapellier\inst{2}\and
	E. Poretti\inst{3} \and
        A. Baglin\inst{4} \and
        W.~W. Weiss\inst{5} \and 
        K. Kolenberg\inst{6,7} \and 
        E. Guggenberger\inst{8,9} \and
        J.-F. Le Borgne\inst{10,11} }
\institute{Konkoly Observatory, MTA CSFK, Konkoly Thege Mikl\'os \'ut 15-17. H-1121 Budapest, Hungary \and 
Laboratoire Lagrange, Universit\'e Nice Sophia-Antipolis, UMR 7293, Observatoire de la C\^ote d'Azur 06300, Nice, France \and 
INAF - Osservatorio Astronomico di Brera, via E. Bianchi 46, 23807 Merate (LC), Italy\and
              LESIA, Universit\'e Pierre et Marie Curie, Universit\'e Denis Diderot, Observatoire de Paris, 92195 Meudon Cedex, France\and 
	     Institute of Astronomy, University of Vienna, T\"urkenschanzstrasse 17, 1180 Vienna, Austria\and 
	     Harvard-Smithsonian Center for Astrophysics, 60 Garden Street, Cambridge MA 02138, USA\and 
Instituut voor Sterrenkunde, K.U. Leuven, Celestijnenlaan 200D, B-3001 Heverlee, Belgium\and 
Max Planck Institute for Solar System Research, Justus-von-Liebig-Weg 3, 37077 G\"ottingen, Germany\and
Stellar Astrophysics Centre, Department of Physics and Astronomy, Aarhus
University, Ny Munkegade 120, 8000 Aarhus C, Denmark\and
Universit\'e de Toulouse, UPS-OMP, IRAP, Toulouse, France\and
CNRS, IRAP, 14, avenue Edouard Belin, F-31400 Toulouse, France}

\abstract{The study of RR\,Lyrae stars has recently been invigorated thanks to the 
long, uninterrupted, ultra-precise time series data provided by the {\it Kepler} and CoRoT space telescopes. We give a brief overview of the new observational 
findings concentrating on the connection between period doubling and the Blazhko modulation, and the omnipresence of additional periodicities in all RR\,Lyrae subtypes, except for non-modulated RRab stars. Recent theoretical results demonstrate that if more than two modes are present in a nonlinear dynamical system such as a high-amplitude RR Lyrae star, the outcome is often an extremely intricate dynamical state. Thus, based on these discoveries, an underlying picture of complex dynamical interactions between modes is emerging which sheds new light on the century-old Blazhko-phenomenon, as well. New directions of theoretical efforts, like multidimensional hydrodynamical simulations, future space photometric missions and detailed spectroscopic investigations will pave the way towards a more complete understanding of the atmospheric and pulsation dynamics of these enigmatic touchstone objects.} 
\maketitle
\section{Introduction}\label{intro}

High precision, uninterrupted, space-based photometric data sets obtained with MOST \cite{walker2003}, CoRoT \cite{baglin2006} and {\it Kepler} \cite{borucki2010} have transformed our view of exoplanetary systems and stellar variability, as well. RR\,Lyrae stars are no exception. 
New types of variations and dynamical phenomena have been discovered, like period doubling \cite{kolenberg2010}, \cite{szabo2010}, the presence of additional pulsational modes \cite{moskalik2014a}, \cite{benko2010}, high-order resonances \cite{kollath2011}, and maybe chaos \cite{plachy2013}. In addition, the mysterious Blazhko effect could be investigated in much greater detail than previously \cite{guggenberger2011}, \cite{guggenberger2012}, \cite{leborgne2014}, \cite{benko2014}. While a great deal of advancement has been achieved on the Blazhko effect itself, the recognition of multiple modulations based on ground-based \cite{sodor2011}, \cite{skarka2014} and space data \cite{benko2014} being one particular example, we refer to a recent summary 
\cite{szabo2014a} regarding the Blazhko effect. In this contribution we provide an update on the most recent results and highlight some emerging trends focusing primarily on period doubling and the presence and dynamics of additional modes in RR\,Lyrae stars. 

\section{Period doubling}\label{sec:1}

Period doubling (PD) is a well-known dynamical phenomenon. Its presence is betrayed by the alternating maxima and light curve shape in the light curve of a pulsating star, while in the frequency spectrum half-integer multiples of the fundamental frequency appear. That means they are located halfway between the dominant pulsation mode and its harmonics \cite{szabo2010}. The importance of PD lies in the fact that transition between regular to chaotic dynamics can occur through a series of PD bifurcations. In RR\,Lyrae stars PD was first discovered in the {\it Kepler} data \cite{kolenberg2010}, \cite{szabo2010}. The origin of the PD could be unambiguously traced back to a 9:2 resonance between the fundamental mode and the ninth radial overtone. In 2011 Buchler \& Koll\'ath \cite{buchler2011} demonstrated that the same resonance may be able to cause light curve modulation, ie. the Blazhko effect. By revisiting the CoRoT RR\,Lyrae light curves we found signs of the PD in four Blazhko RRab stars \cite{szabo2014b} out of a sample of six. A recent comprehensive work by Benk\H o et al. \cite{benko2014} on {\it Kepler} modulated RRab stars gave concordant results: six out of ten Blazhko-modulated RRab stars exhibited PD. Based on the latest space photometric results we can conclude that altogether two-thirds of the Blazhko-modulated RRab stars exhibit this new type of dynamical phenomenon \cite{szabo2014b}.  It is important to emphasize that PD in connection with the dominant pulsation mode has been only seen in Blazhko-modulated RRab stars. No non-modulated RR\,Lyrae show PD down to the outstanding precision delivered by CoRoT and {\it Kepler}.

\begin{figure*}
\resizebox{1.00\columnwidth}{!}{\includegraphics{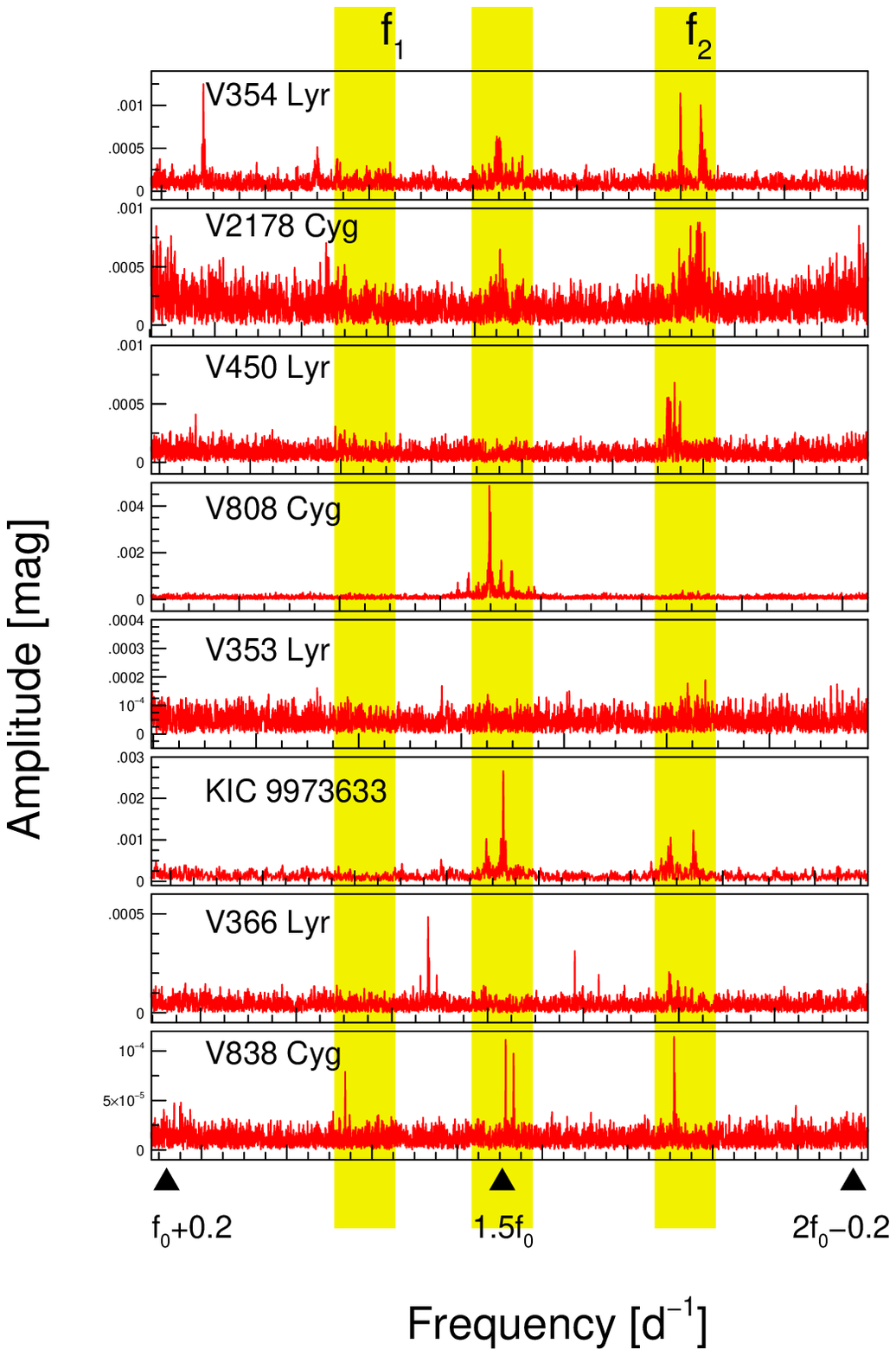} \includegraphics{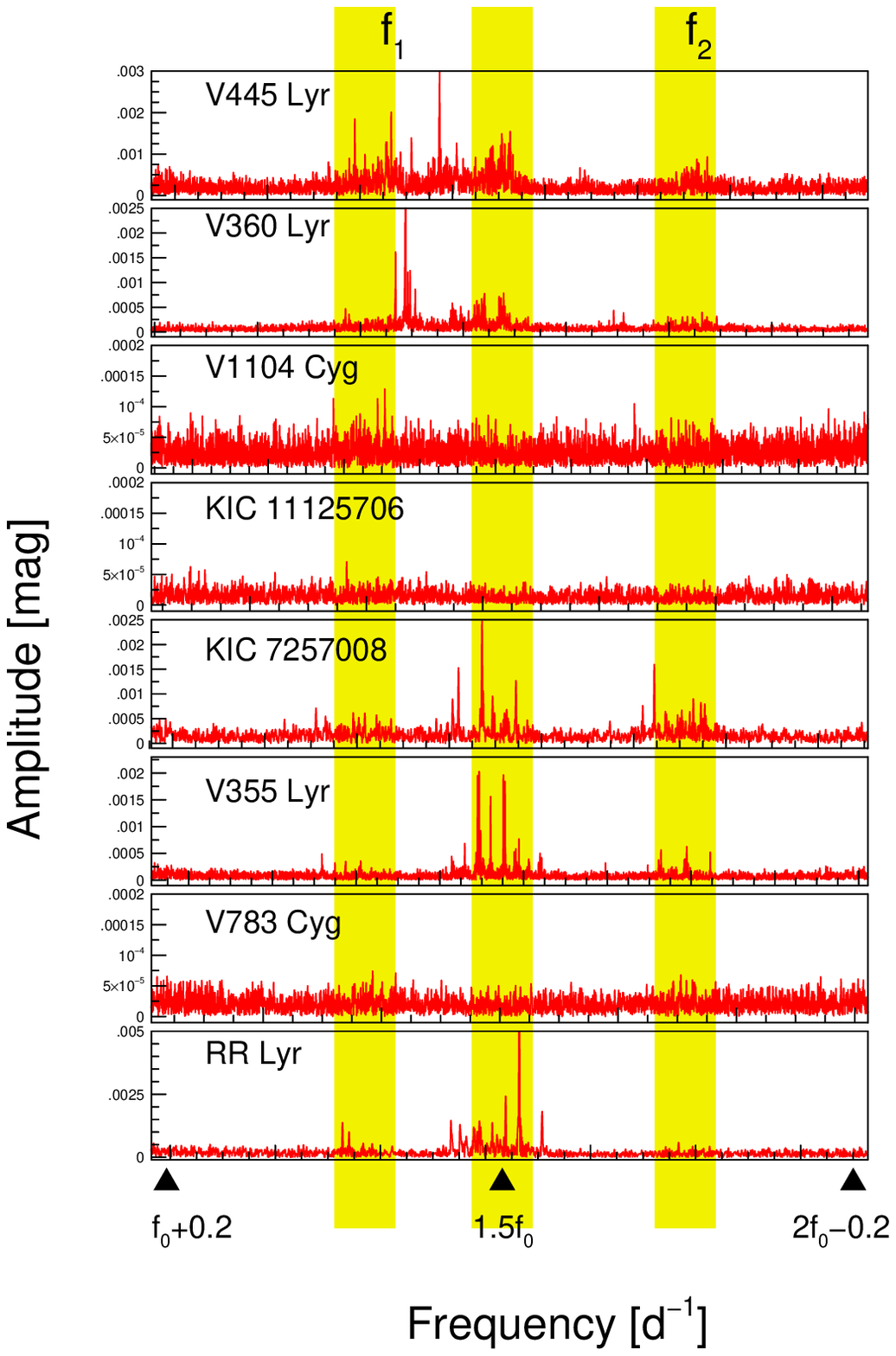}}
\caption{Additional frequencies in {\it Kepler} Blazhko RRab stars \cite{kolenberg2011}, \cite{benko2014}. The frequency interval between the fundamental mode and its first harmonic is plotted. From left to right the yellow stripes show  the location of the first radial overtone, the half-integer frequencies caused by period doubling, and the second radial overtones. The significant frequencies around the second overtone in V808\,Cyg are barely visible because of the large amplitude of the half-integer frequency, see Fig. 11 in \cite{benko2014} for a close-up view.}
\label{fig:1}      
\end{figure*}

\section{Additional modes}\label{sec:2}

\begin{figure*}
\resizebox{1.00\columnwidth}{!}{\includegraphics{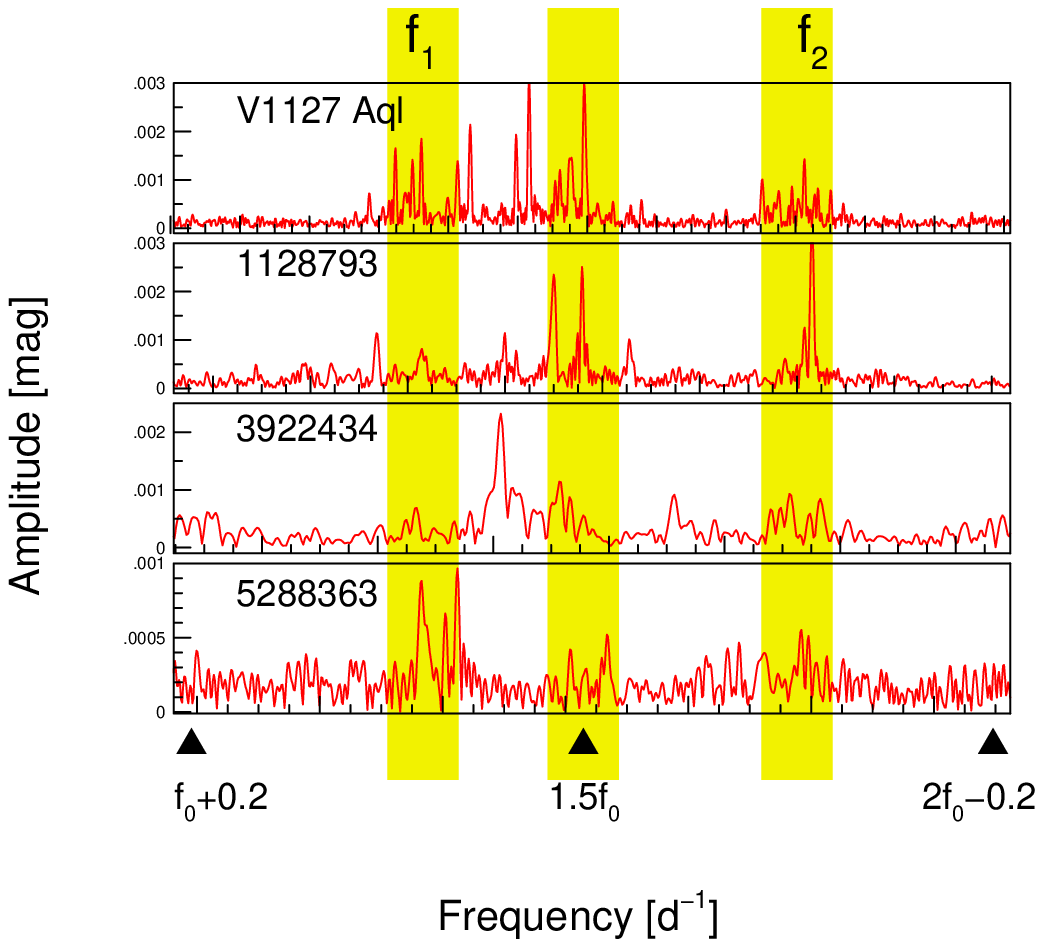}
\includegraphics{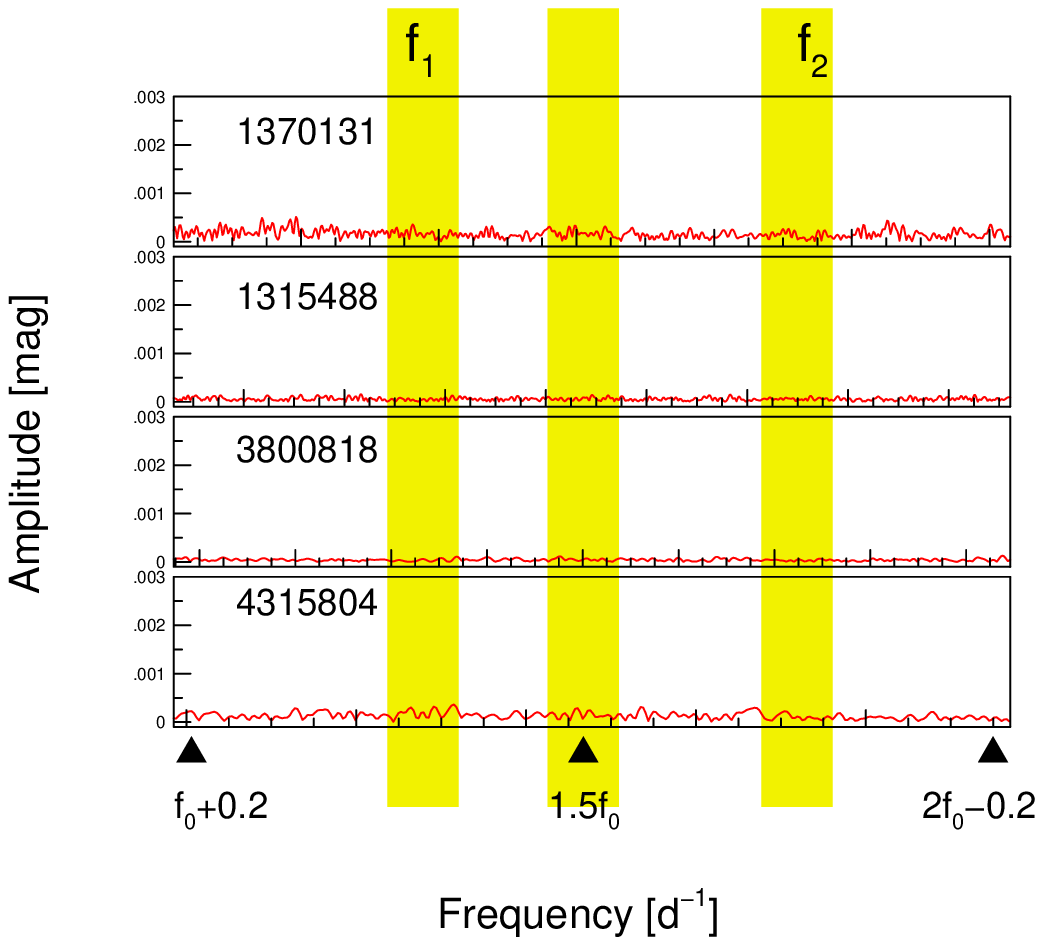}}
\caption{Additional frequencies in CoRoT RRab stars. Left panel: Blazhko-modulated stars, right panel: non-Blazhko stars. The notation is the same as in Fig~\ref{fig:1}.  The lack of any additional frequencies in non-modulated stars is remarkable.}
\label{fig:2}      
\end{figure*}

RR\,Lyrae stars have been known for decades to pulsate exclusively in the fundamental (F) or first overtone (O1) modes, or occasionally in both (RRd, double-mode stars). With the advent of high quality space data it became obvious that most of the RR\,Lyrae stars show additional periodicities beyond these dominant, low-order radial modes.
In modulated RRab stars we find power in the frequency spectrum around the theoretical 
location of the first and/or second radial overtone (see Fig~\ref{fig:1} for the {\it Kepler} sample \cite{benko2010}, \cite{benko2014} and the left panel of Fig~\ref{fig:2} for the CoRoT Blazhko stars \cite{szabo2014b}). These can be either the radial overtones themselves as was demonstrated by Moln\'ar et al. \cite{molnar2012}, or nonradial modes close to or exactly in 1:1 resonance with the radial modes predicted by earlier theoretical works \cite{vanhoolst1998} \cite{dziembowski2004}. Interestingly, 
none of the non-modulated RRab stars exhibit additional frequencies. This dichotomy is strikingly demonstrated in the right panel of Fig~\ref{fig:2}. Overtone pulsators (RRc stars) also show additional frequencies, most prominent and most frequent of them is a 
probably nonradial pulsation mode with a period ratio of $\sim$0.61 with respect to the dominant overtone radial mode. Interestingly, it is present in almost all RRc and RRd stars, ie. in those where the dominant mode is the first radial overtone \cite{szabo2014b}. A detailed summary on the topic is recently given by Moskalik et al. \cite{moskalik2014b}. Our conclusion is that additional modes are universal in RR Lyrae except for non-modulated RRab stars. This finding has ramifications for the nature of the mysterious Blazhko effect. In summary, Blazhko-modulation, period doubling and additional modes seem to be related phenomena, all being part of an intricate and complex underlying dynamics. 

We investigated the temporal variability of the additional modes in the CoRoT RR Lyrae sample \cite{szabo2014b}. It turned out that the amplitude or the shape of these frequencies vary in time in most cases where we could draw firm conclusions. This variability has been confirmed by Moskalik et al. \cite{moskalik2014b} based on the {\it Kepler} RRc sample. The {\it Kepler} data provide better frequency resolution compared to our CoRoT sample, hence the presence of potentially close-by, unresolved frequency components could be safely excluded as the main cause of the variations. Half-integer frequencies are nonstationary, because of the ephemeral presence of PD itself. However, the variability of the other (presumably) nonradial modes is more intriguing. While in Blazhko RRab stars the temporal changes may be related to the modulation mechanism (though more work should be done to corroborate this connection), in non-modulated RR\,Lyrae, such as RRd and RRc stars, however, a different mechanism may be at work. The most probably culprit  is an intricate dynamical interaction between the radial and nonradial modes. The most important observational result based on the growing number 
of RR\,Lyrae stars observed from space is that temporal variability of the additional modes seems to be ubiquitous in these objects. 

\section{Outlook}\label{sec:4}

Space photometry induced a veritable revolution in our understanding of RR Lyrae stars. New dynamical phenomena have been discovered, of which some are well-understood \cite{kollath2011}, while others are still awaiting a theoretical explanation. If the ubiquity of nonradial modes is confirmed, the power of nonlinear seismology using radial modes in RR Lyrae stars can be unleashed \cite{molnar2012}. We hope to continue this adventure using a larger sample of galactic and maybe extragalactic RR\,Lyrae stars by ongoing and future space photometric missions, like K2 \cite{howell2014}, \cite{molnar2014}, TESS \cite{ricker2014}, and PLATO \cite{rauer2014}.

\begin{acknowledgement}
{\bf Acknowledgements.} This project has been supported by the 
Hungarian OTKA grant K83790 and the European Community's Seventh 
Framework Program (FP7/2007-2013) under grant agreements no. 312844 
(SPACEINN), no. 269194 (IRSES/ASK) and ERC grant agreement no. 338251 (StellarAges).  RSz, MP, and JMB acknowledge the support of the ESA PECS project 
No.~4000103541/11/NL/KML. WW was supported by the Austrian Science Fonds (FWF P22691-N16). KK acknowledges the support of FP7  Marie Curie Fellowship PIOF-255267 (SAS-RRL). This research made use of the ExoDat database, operated at LAM-OAMP, Marseille, France, on behalf of the CoRoT/Exoplanet program. Funding for the {\it Kepler} mission is provided by the NASA Science Mission directorate. We are grateful to the Kepler Science Team for their extensive efforts in producing and making publicly available these high-quality and unique data sets.
\end{acknowledgement}


\begin{thebibliography}{}
\bibitem{baglin2006} Baglin, A., Auvergne, M., Boisnard, L., et al. in {\it 36th COSPAR Scientific Assembly}, \textbf{36}, 3749 (2006)
\bibitem{benko2010} Benk\H o, J. M., Kolenberg, K., Szab\'o, R., et al., MNRAS \textbf{409}, 1585 (2010)
\bibitem{benko2014} Benk\H o, J. M., Plachy, E., Szab\'o, R., et al., ApJS \textbf{213}, 31 (2014)
\bibitem{borucki2010} Borucki, W. J., Koch, D., Basri, G., et al., Science \textbf{327}, 977 (2010)
\bibitem{buchler2011} Buchler, J. R., Koll\'ath, Z. ApJ \textbf{731}, 24 (2011)
\bibitem{dziembowski2004} Dziembowski, W. A., Mizerski, T., AcA 54, \textbf{363} (2004)
\bibitem{guggenberger2011} Guggenberger, E., Kolenberg, K., Chapellier, E., et al. MNRAS \textbf{415}, 1577 (2011) 
\bibitem{guggenberger2012} Guggenberger, E., Kolenberg, K., Nemec, J. M., et al. MNRAS \textbf{424}, 649 (2012)
\bibitem{howell2014} Howell, S. B., Sobeck, C., Haas, M., et al., PASP, \textbf{126}, 398 (2014)
\bibitem{kolenberg2010} Kolenberg, K., Szab\'o, R., Kurtz, D. W., et al., ApJ \textbf{713}, L198 (2010)
\bibitem{kolenberg2011} Kolenberg, K., Bryson, S. T., Szab\'o, R., et al., MNRAS  \textbf{411}, 878 (2011)
\bibitem{kollath2011} Koll\'ath, Z., Moln\'ar, L., Szab\'o, R. MNRAS \textbf{414}, 1111 (2011)
\bibitem{leborgne2014} Le Borgne, J. F., Poretti, E., Klotz, A., et al., MNRAS, \textbf{441}, 1435 (2014)
\bibitem{moskalik2014a} Moskalik, P., Proc. of IAU Symp. \textbf{301}, 249 (2014a)
\bibitem{molnar2012} Moln\'ar, L., Koll\'ath, Z., Szab\'o, R., et al., ApJL \textbf{757}, L13, (2012)
\bibitem{molnar2014} Moln\'ar, L., Plachy, E., Szab\'o, R., IBVS \textbf{6108} (2014)
\bibitem{moskalik2014b} Moskalik, P., Smolec, R., Kolenberg, K., et al., MNRAS submitted (2014b)
\bibitem{plachy2013} Plachy, E., Koll\'ath, Z., Moln\'ar, L., MNRAS \textbf{433}, 3590 (2013)
\bibitem{rauer2014} Rauer, H., Catala, C., Aerts, C., et al. Exp. Astron. in press (2014)
\bibitem{ricker2014} Ricker, G. R., Winn, J. N., Vanderspek, R., et al., Proc. of the SPIE, \textbf{9143}, 15 (2014)
\bibitem{skarka2014} Skarka, M. A\&A \textbf{562}, A90 (2014)
\bibitem{sodor2011} S\'odor, \'A., Jurcsik, J., Szeidl, B. et al., MNRAS, \textbf{411}, 1585 (2011)
\bibitem{szabo2010} Szab\'o, R., Koll\'ath, Z., Moln\'ar, L., et al., MNRAS \textbf{409}, 1244 (2010)
\bibitem{szabo2014a} Szab\'o, R., Proc. of IAU Symp. \textbf{301}, 241 (2014a) 
\bibitem{szabo2014b} Szab\'o, R., Benk\H o, J. M., Papar\'o, M., et al., A\&A \textbf{570}, A100, (2014b) 
\bibitem{vanhoolst1998} Van Hoolst, T., Dziembowski, W. A., Kawaler, S. D., MNRAS, \textbf{297}, 536 (1998)
\bibitem{walker2003} Walker, G., Matthews, J., Kuschnig, R., et al., PASP  \textbf{115}, 1023 (2003)
\end{thebibliography}
\end{document}